\documentclass[prl,aps,twocolumn,superscriptaddress,showpacs,showkeys,amsmath,amssymb,floatfix, longbibliography]{revtex4-1}
\usepackage[colorlinks=true,citecolor=blue,filecolor=blue,linkcolor=blue,urlcolor=blue,pdftex]{hyperref}
\usepackage{currfile} 
\usepackage{soul} 
\usepackage[separate-uncertainty = true, multi-part-units=single]{siunitx}
\usepackage{graphicx}
\usepackage{dcolumn}
\usepackage{bm}
\usepackage[section]{placeins} 
\usepackage{upgreek}
\usepackage{isotope}
\usepackage[mathlines]{lineno}
\newcommand*\patchAmsMathEnvironmentForLineno[1]{%
  \expandafter\let\csname old#1\expandafter\endcsname\csname #1\endcsname
  \expandafter\let\csname oldend#1\expandafter\endcsname\csname end#1\endcsname
  \renewenvironment{#1}%
     {\linenomath\csname old#1\endcsname}%
     {\csname oldend#1\endcsname\endlinenomath}}%
\newcommand*\patchBothAmsMathEnvironmentsForLineno[1]{%
  \patchAmsMathEnvironmentForLineno{#1}%
  \patchAmsMathEnvironmentForLineno{#1*}}%
\AtBeginDocument{%
\patchBothAmsMathEnvironmentsForLineno{equation}%
\patchBothAmsMathEnvironmentsForLineno{align}%
\patchBothAmsMathEnvironmentsForLineno{flalign}%
\patchBothAmsMathEnvironmentsForLineno{alignat}%
\patchBothAmsMathEnvironmentsForLineno{gather}%
\patchBothAmsMathEnvironmentsForLineno{multline}%
}

\newcommand{\bologna}{\affiliation{Department of Physics and Astronomy, University of Bologna and INFN-Bologna, 40126 Bologna, Italy}}

\newcommand{\chicago}{\affiliation{Department of Physics \& Kavli Institute for Cosmological Physics, University of Chicago, Chicago, IL 60637, USA}}

\newcommand{\coimbra}{\affiliation{LIBPhys, Department of Physics, University of Coimbra, 3004-516 Coimbra, Portugal}}

\newcommand{\columbia}{\affiliation{Physics Department, Columbia University, New York, NY 10027, USA}}

\newcommand{\lngs}{\affiliation{INFN-Laboratori Nazionali del Gran Sasso and Gran Sasso Science Institute, 67100 L'Aquila, Italy}}

\newcommand{\mainz}{\affiliation{Institut f\"ur Physik \& Exzellenzcluster PRISMA, Johannes Gutenberg-Universit\"at Mainz, 55099 Mainz, Germany}}

\newcommand{\heidelberg}{\affiliation{Max-Planck-Institut f\"ur Kernphysik, 69117 Heidelberg, Germany}}

\newcommand{\munster}{\affiliation{Institut f\"ur Kernphysik, Westf\"alische Wilhelms-Universit\"at M\"unster, 48149 M\"unster, Germany}}

\newcommand{\nikhef}{\affiliation{Nikhef and the University of Amsterdam, Science Park, 1098XG Amsterdam, Netherlands}}

\newcommand{\nyuad}{\affiliation{New York University Abu Dhabi, Abu Dhabi, United Arab Emirates}}

\newcommand{\purdue}{\affiliation{Department of Physics and Astronomy, Purdue University, West Lafayette, IN 47907, USA}}

\newcommand{\rpi}{\affiliation{Department of Physics, Applied Physics and Astronomy, Rensselaer Polytechnic Institute, Troy, NY 12180, USA}}

\newcommand{\rice}{\affiliation{Department of Physics and Astronomy, Rice University, Houston, TX 77005, USA}}

\newcommand{\stockholm}{\affiliation{Oskar Klein Centre, Department of Physics, Stockholm University, AlbaNova, Stockholm SE-10691, Sweden}}

\newcommand{\subatech}{\affiliation{SUBATECH, IMT Atlantique, CNRS/IN2P3, Universit\'e de Nantes, Nantes 44307, France}}

\newcommand{\torino}{\affiliation{INFN-Torino and Osservatorio Astrofisico di Torino, 10125 Torino, Italy}}

\newcommand{\ucla}{\affiliation{Physics \& Astronomy Department, University of California, Los Angeles, CA 90095, USA}}

\newcommand{\ucsd}{\affiliation{Department of Physics, University of California, San Diego, CA 92093, USA}}

\newcommand{\wis}{\affiliation{Department of Particle Physics and Astrophysics, Weizmann Institute of Science, Rehovot 7610001, Israel}}

\newcommand{\zurich}{\affiliation{Physik-Institut, University of Zurich, 8057  Zurich, Switzerland}}

\newcommand{\paris}{\affiliation{LPNHE, Sorbonne Universit\'{e}, Universit\'{e} Paris Diderot, CNRS/IN2P3, Paris 75252, France}}

\newcommand{\freiburg}{\affiliation{Physikalisches Institut, Universit\"at Freiburg, 79104 Freiburg, Germany}}

\newcommand{\lal}{\affiliation{LAL, Universit\'e Paris-Sud, CNRS/IN2P3, Universit\'e Paris-Saclay, F-91405 Orsay, France}}

\newcommand{\naples}{\affiliation{Department of Physics ``Ettore Pancini'', University of Napoli and INFN-Napoli, 80126 Napoli, Italy}} 

\begin{document}

\title{First observation of two-neutrino double electron capture in \isotope[124]{Xe} with XENON1T}

\author{E.~Aprile}\columbia
\author{J.~Aalbers}\stockholm\nikhef
\author{F.~Agostini}\bologna
\author{M.~Alfonsi}\mainz
\author{L.~Althueser}\munster
\author{F.~D.~Amaro}\coimbra
\author{M.~Anthony}\columbia
\author{V.~C.~Antochi}\stockholm
\author{F.~Arneodo}\nyuad
\author{L.~Baudis}\zurich
% ...
\collaboration{XENON Collaboration}
\email[]{xenon@lngs.infn.it}
\noaffiliation

\date{December 20, 2018 }




\begin{abstract}
Two-neutrino double electron capture ($2\upnu$ECEC) is a second-order Weak process with predicted half-lives that surpass the age of the Universe by many orders of magnitude \cite{Winter:1955zz}. Until now, indications for $2\upnu$ECEC decays have only been seen for two isotopes, \isotope[78]{Kr} \cite{PhysRevC.87.035501, PhysRevC.96.065502} and \isotope[130]{Ba} \cite{PhysRevC.64.035205, PUJOL20096834}, and instruments with very low background levels are needed to detect them directly with high statistical significance \cite{Gavriljuk2018, Abe:2018gyq}. The $2\upnu$ECEC half-life provides an important input for nuclear structure models \cite{Pirinen:2015sma,Suhonen:2013rca,Aunola:1996ui,Singh:2007jh,Hirsch1994,Rumyantsev:1998uy,Perez:2018cly} and its measurement represents a first step in the search for the neutrinoless double electron capture processes ($0\upnu$ECEC). A detection of the latter would have implications for the nature of the neutrino and give access to the absolute neutrino mass \cite{Majorana1937, Bernabeu:1983yb, Sujkowski:2003mb}. Here we report on the first direct observation of $2\upnu$ECEC in \isotope[124]{Xe} with the XENON1T Dark Matter detector. The significance of the signal is $4.4\sigma$ and the corresponding half-life $T_{1/2}^{2\upnu\text{ECEC}} = (1.8\pm 0.5_\text{stat}\pm 0.1_\text{sys})\times 10^{22}\;\text{y}$ is the longest ever measured directly. This study demonstrates that the low background and large target mass of xenon-based Dark Matter detectors make them well suited to measuring other rare processes as well, and it highlights the broad physics reach for even larger next-generation experiments \cite{aprile15c, Mount:2017qzi, Aalbers:2016jon}.
\end{abstract}

\maketitle

\begin{figure}[b]
    \centering
    \includegraphics[width=0.45\textwidth]{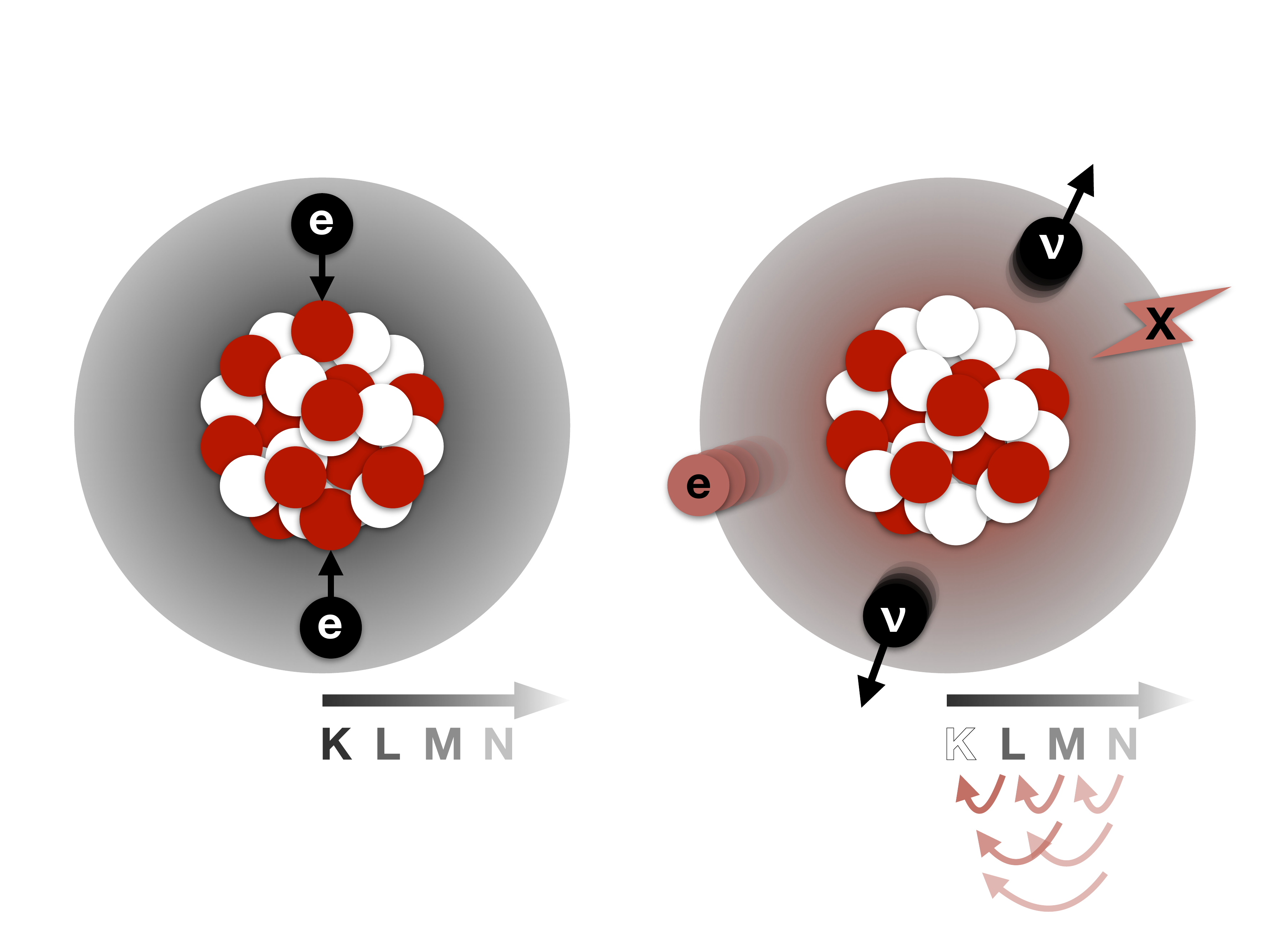}
    \caption{In the $2\upnu$ECEC process the nucleus captures two atomic shell electrons (black), most likely from the K-shell, and simultaneously converts two protons (red) to neutrons (white). Two neutrinos (black) are emitted in the nuclear process and carry away most of the decay energy while the atomic shell is left in an excited state with two holes in the K-shell. A cascade of X-rays (red X) and Auger electrons (red e) are emitted in the atomic relaxation where the lower shells are refilled from the higher ones (arrows).}
    \label{fig:dec_schematic}
\end{figure}

\begin{figure}
    \centering
    \includegraphics[width=0.45\textwidth]{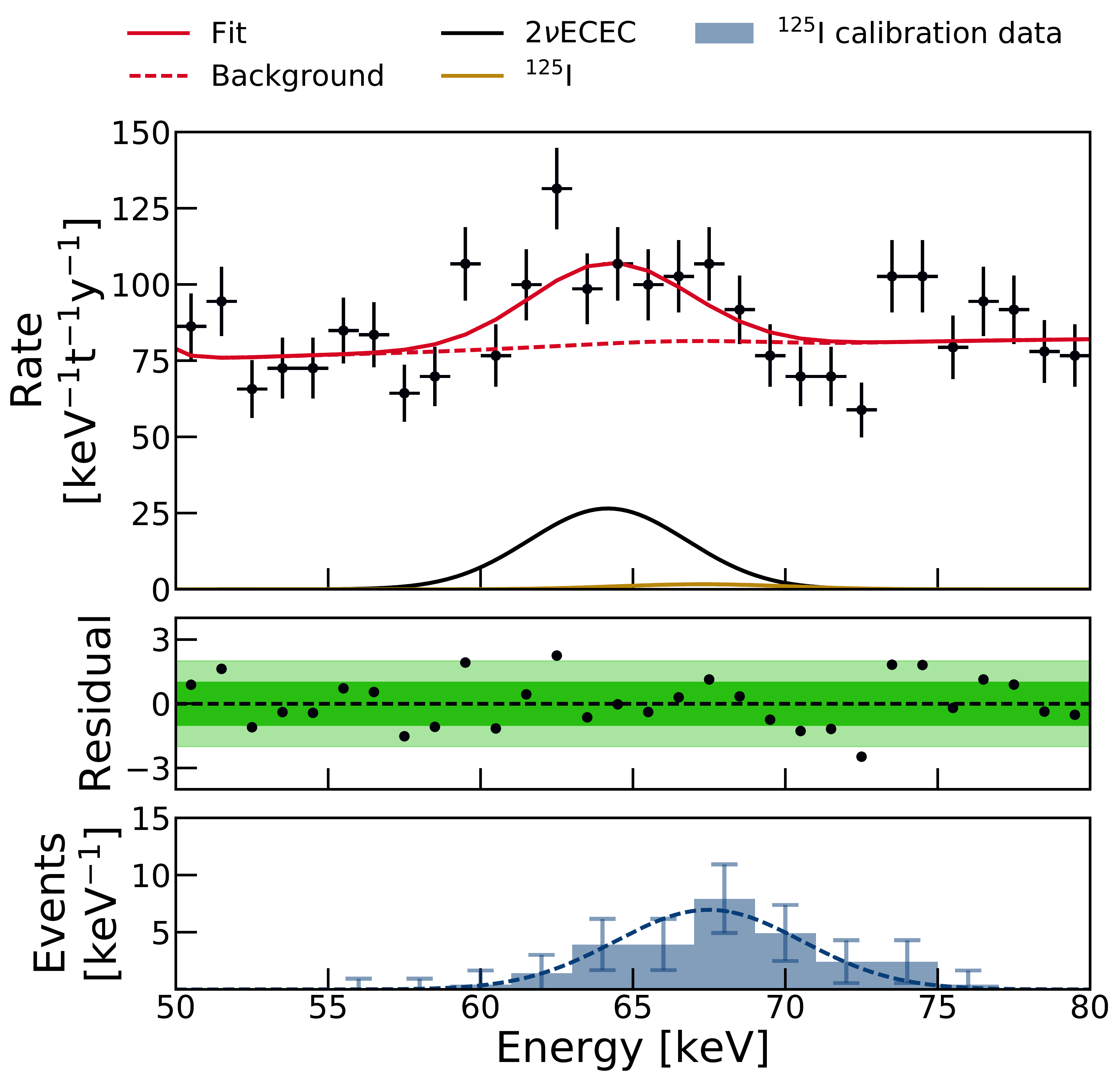}
    \caption{Zoom on the energy region of interest for 2$\upnu$ECEC in $^{124}$Xe. The best fit contribution from $2\upnu$ECEC with $N_\text{$2\upnu$ECEC}=126$ events is given by the solid black line while the full fit is indicated as the solid red line. The peak from $^{125}$I with $N_\text{\isotope[125]{I}} = 9$ events is indicated by the solid gold line. The background-only model without $2\upnu$ECEC (red dashed) clearly does not describe the data. Residuals for the best fit are given in the central panel with the 1$\sigma$\,(2$\sigma$) band indicated in green\,(light green). The bottom panel shows a histogram of the \isotope[125]{I} activation peak as seen in \SI{6}{\day} of data after a dedicated neutron generator calibration. A linear background has been subtracted from the data and the peak shows the expected shift with respect to the $2\upnu$ECEC signal.}
    \label{fig:dec_fit}
\end{figure}

\indent The long half-life of double electron capture makes it extremely rare and the process has escaped detection for decades.~In the two-neutrino case ($2\upnu$ECEC), two protons in a nucleus simultaneously convert into neutrons by the absorption of two electrons from one of the atomic shells and the emission of two electron neutrinos ($\upnu_e$) \cite{Winter:1955zz}. After the capture of the two atomic electrons, mostly from the K shell \cite{Doi:1992dm}, the filling of the vacancies results in a detectable cascade of X-rays and Auger electrons \cite{relax}. The nuclear binding energy $Q$ released in the process ($\mathcal{O}$(MeV)) is carried away by the two neutrinos, which are not detected within the detector. Thus, the experimental signature appears in the keV-range rather than the MeV-range. The process is illustrated in Fig.~\ref{fig:dec_schematic}.
\\
\indent $2\upnu$ECEC is allowed in the Standard Model of particle physics and related to double $\upbeta$-decay as a second-order Weak Interaction process. However, few experimental indications exist. Geochemical studies for \isotope[130]{Ba} \cite{PhysRevC.64.035205, PUJOL20096834} and a direct measurement for \isotope[78]{Kr} \cite{PhysRevC.87.035501, PhysRevC.96.065502} quote half-lives on the order of $10^{20}-10^{22}$ years.\\ 
\indent Even longer timescales are expected for a hypothetical double electron capture without neutrino emission ($0\upnu$ECEC) \cite{Bernabeu:1983yb, Sujkowski:2003mb}. A detection of this decay would show that neutrinos are Majorana particles \cite{Majorana1937}, i.e. their own anti-particles, and could help understanding the dominance of matter over antimatter in our Universe by means of Leptogenesis \cite{buchmueller05leptogenesis}. An eventual Majorana nature would give access to the absolute neutrino mass, but rely on nuclear matrix element calculations from theory. A plethora of different calculation approaches and results exist \cite{Pirinen:2015sma,Suhonen:2013rca,Aunola:1996ui,Singh:2007jh,Hirsch1994,Rumyantsev:1998uy,Perez:2018cly}. As these models also predict the $2\upnu$ECEC half-life, its measurement would provide necessary input to narrow down the uncertainty therein.\\
\indent Here we study the $2\upnu$ECEC of \isotope[124]{Xe}. Natural xenon is a radiopure and scalable detector medium that contains about \SI{1}{\kg} of \isotope[124]{Xe} per tonne. \isotope[124]{Xe} undergoes $2\upnu$ECEC to $^{124}$Te with $Q= \SI{2857}{\keV}$ \cite{Nesterenko:2012xp}. Since the amount of energy released by the recoiling nucleus is negligible ($\mathcal{O}(\SI{10}{\eV})$) and with the neutrinos carrying away the energy $Q$ undetected, only the X-rays and Auger electrons are measured. The total energy for the double K-shell capture is \SI{64.3}{\keV} \cite{Nesterenko:2012xp}. This value has already been corrected for energy depositions that do not exceed the xenon excitation threshold \cite{relax, Szydagis:2011tk}. Previous searches for the $2\upnu$ECEC decay of \isotope[124]{Xe} have been carried out with gas proportional counters using enriched xenon \cite{Gavriljuk2018} as well as large detectors originally designed for Dark Matter searches \cite{Aprile:2016qsw}. The currently leading lower limit on the half-life comes from the XMASS collaboration at $T_{1/2}^{2\upnu\text{ECEC}} > 2.1 \times 10^{22}\;\text{y}$ ($90\,\%$ C.L.) \cite{Abe:2018gyq}.\\
\indent XENON1T \cite{Aprile:2017aty} was built to detect interactions of Dark Matter in the form of weakly interacting massive particles (WIMPs) and has recently placed the most stringent limits on the coherent elastic scattering of WIMPs with xenon nuclei \cite{Aprile:2018prl}. XENON1T uses \SI{3.2}{t} of ultra-pure liquid xenon (LXe), of which \SI{2}{t} are within the sensitive volume of the time projection chamber (TPC): a cylinder of $\sim$\SI{96}{\centi\meter} diameter and height with walls of highly-reflective PTFE that is instrumented with 248 photomultiplier tubes (PMTs). The TPC allows for the measurement of the scintillation (S1) and ionisation signals (S2) induced by a particle interaction -- the latter by converting ionisation electrons into light by means of proportional scintillation. It provides calorimetry, 3D position reconstruction, and measures the scatter multiplicity.\\
\indent The detector is shielded by the overburden due to its underground location at Laboratori Nazionali del Gran Sasso, an active water Cherenkov muon veto \cite{Aprile:2014jin}, and the liquid xenon itself. All detector materials were selected for low amounts of radioactive impurities and low radon emanation rates \cite{Aprile:2017scr}. In addition, the anthropogenic $\upbeta$-emitter \isotope[85]{Kr} was removed from the xenon inventory by cryogenic distillation \cite{Aprile:2017kry}.~The combination of material selection, active background reduction, and an inner low-background fiducial volume selection in data analysis results in an extremely low event rate. This makes XENON1T the currently most sensitive detector for $2\upnu$ECEC searches in \isotope[124]{Xe}.\\
\indent The data presented here was recorded between February 2, 2017 and February 8, 2018 as part of a Dark Matter search. Details on the detector conditions and signal corrections can be found in the original publication \cite{Aprile:2018prl}. The data quality criteria from the Dark Matter analysis were applied with the exception of those exhibiting low acceptance in the energy region of interest around \SI{60}{\keV}. During the analysis, the data was blinded, i.e. inaccessible for analysis, from \SI{56}{\keV} to \SI{72}{\keV} and only unblinded after the data quality criteria, fiducial volume, and background model had been fixed. Data sets acquired after detector calibrations with an external $^{241}$AmBe neutron source or a deuterium-deuterium-fusion neutron generator were removed in order to reduce the impact of radioactive \isotope[125]{I}. It is produced by the activation of \isotope[124]{Xe} during neutron calibrations and is taken out within a few days through the purification system. A pre-unblinding quantification of this removal using short-term calibration data led to a first reduction of the data set to \SI{214.3}{days}. This data was used for fixing the background model. After unblinding, the long-term behaviour of \isotope[125]{I} could be quantified and led to a further removal of data sets (methods). This yielded a final live time of \SI{177.7}{days}.\\
\indent Atomic X-rays and Auger electrons cannot be resolved individually due to their sub-millimetre range in LXe and the fast atomic processes. Thus, the experimental signature of K-shell $2\upnu$ECEC in XENON1T is a single S1~+~S2 pair. Both S1 and S2 signals are used for the analysis to achieve the optimal energy resolution \cite{conti2003} for the resulting peak. The energy scale around the expected signal at $E_0 = (64.3\pm 0.6)\;\text{keV}$ is calibrated using mono-energetic lines of injected calibration sources (e.g.\isotope[83\text{m}]{Kr}), neutron-activated xenon isotopes, and $\upgamma$-rays from radioactive decays in detector materials. The energy resolution of a Gaussian peak at $E_0$ is $\sigma/\mu = (4.1\pm 0.4)\,\%$ (methods). The uncertainty on $E_0$ reflects the uncertainties of both the energy reconstruction and the correction for sub-excitation quanta. An ellipsoidal \SI{1.5}{\tonne} inner fiducial mass was identified as providing the optimal signal-to-background ratio in sideband studies between \SI{80}{\keV} and \SI{140}{\keV}, above the blinded signal region.
\begin{figure}
    \centering
    \includegraphics[width=0.45\textwidth]{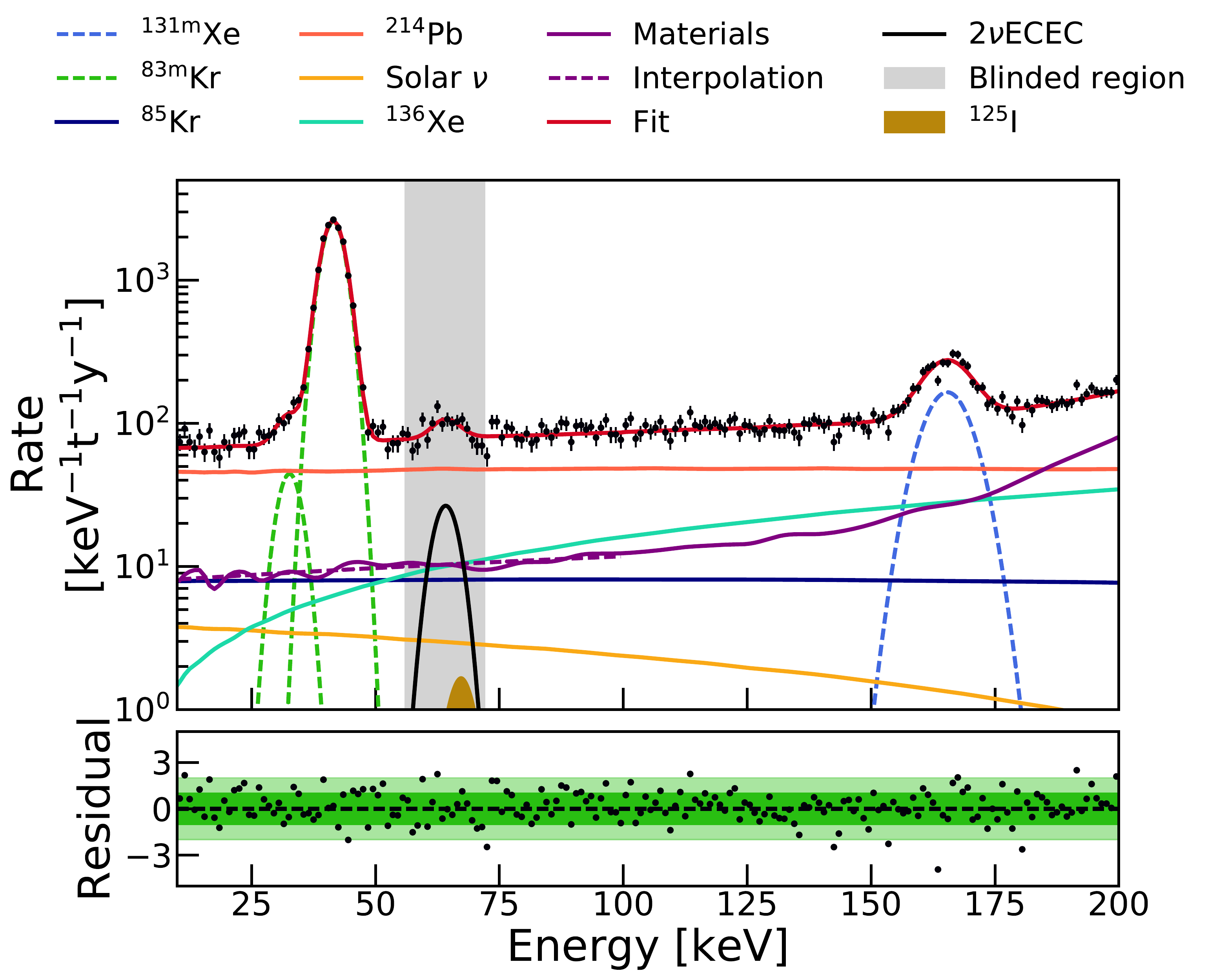}
    \caption{Measured background energy spectrum in the 1.5~t inner fiducial mass, in which the signal-to-background ratio was found to be optimal. The data is described by a simultaneous fit of Monte Carlo generated background spectra, taking into account all known background sources and the $2\upnu$ECEC signal (solid red line, $\chi^2$/d.o.f. $\approx$ 527.3/462). The linear interpolation of material backgrounds below 100~keV is indicated as the purple dashed line. The energy region around the expected $2\upnu$ECEC peak was blinded (grey band) until the background model was defined. The lower panel shows the residuals between the data and the fit including  1$\sigma$\,(2$\sigma$) bands in green\,(light green).} 
    \label{fig:sideband_fit}
\end{figure}\\
\indent Understanding the measured energy spectrum is essential when searching for a small peak from $2\upnu$ECEC. Three classes of backgrounds contribute to the spectrum: intrinsic radioactive isotopes that are mixed with the LXe, radioactive isotopes in the detector materials, and solar neutrinos. The latter is subdominant and well-constrained from solar and nuclear physics. $\upgamma$-rays from \isotope[60]{Co}, \isotope[40]{K}, as well as from \isotope[238]{U} and \isotope[232]{Th} decay chains constitute the bulk of the material backgrounds. They can undergo forward Compton scattering before entering the \SI{2.0}{\tonne} active mass and produce a flat spectrum at low energies. Multiple scatters inside the active volume are rejected by selecting events with only a single S2 compatible with a single S1. The most important intrinsic background components are $\upbeta$-decays of \isotope[214]{Pb}, a daughter of \isotope[222]{Rn} that is emanated from inner surfaces in contact with xenon, the two-neutrino double $\upbeta$-decay of \isotope[136]{Xe}, and the $\upbeta$-decay of \isotope[85]{Kr}. Mono-energetic peaks from \isotope[83\text{m}]{Kr} injected for calibration and activation peaks that occur after neutron calibrations (\isotope[131\text{m}]{Xe} and \isotope[129\text{m}]{Xe}) are present in the spectrum as well. The activation $\isotope[124]{Xe} + \text{n} \rightarrow \isotope[125]{Xe} + \upgamma$ has implications for $2\upnu$ECEC search as \isotope[125]{Xe} decays to \isotope[125]{I} via electron capture. With a branching ratio of $100\,\%$ and a half-life of $59.4\;\text{d}$, \isotope[125]{I} decays into an excited state of \isotope[125]{Te}. The subsequently emitted $\upgamma$-ray together with the K-shell X-ray, which is produced in $87.5\,\%$ of all cases, leads to a mono-energetic peak at \SI{67.3}{\keV}. Due to its proximity to $E_0$ it would present a major background for the $2\upnu$ECEC search that would only become apparent after unblinding. Using an activation model based on the parent isotope, we verified that \isotope[125]{I} is removed from the detector with a time constant of $\tau=(9.1\pm 2.6)\;\text{d}$ (methods). This is in accordance with the continuous xenon purification using hot zirconium getters \cite{Aprile:2017aty}. Accounting for artificial neutron activation from calibrations and for activation by radiogenic thermal neutrons in the purification loop outside the water tank, we expect $N_{\isotope[125]{I}} = (10\pm 7)\;\text{events}$ in the full data set.\\
\indent The background model was constructed by matching Monte Carlo (MC) simulations of all known background components \cite{aprile15c} with the measured energy spectrum. Taking into account the finite detector resolution, events with single energy depositions in the active volume were selected from the MC data and convolved with the measured energy resolution. The weighted sum of all spectra was optimised simultaneously to resemble the measured energy spectrum (methods). The blinded signal region was not used in the fit. The measured energy spectrum with the best fits for the individual components is shown in Fig.~\ref{fig:sideband_fit}. After unblinding of the signal region a clear peak at $E_0$ was identified. The energy and signal width obtained from the spectral fit to the unblinded data are $\mu=(64.2\pm 0.5)\;\si{\keV}$ and $\sigma =(2.6\pm 0.3)\;\si{\keV}$, respectively. The resulting sum spectrum of the event rate is shown in Fig.~\ref{fig:dec_fit}. Converting the fit to a total event count yields $N_\text{\isotope[125]{I}} = (9 \pm 7)$ events from the decay of \isotope[125]{I} and $N_\text{$2\upnu$ECEC}=(126\pm 29)$ events from $2\upnu$ECEC. Compared to the null hypothesis the $\sqrt{\Delta\chi^2}$ of the best-fit is 4.4.\\ 
\indent Several consistency checks have been carried out. It was verified that the signal is homogeneously distributed in space and we checked that the signal accumulates linearly with the exposure. A simultaneous fit of an inner (\SI{1.0}{t}) and outer (\SI{0.5}{t}) detector mass with different background compositions yielded consistent signal rates. We verified the linearity of the energy calibration by identifying the \isotope[125]{I} activation peak at its expected position, which is separated from $E_0$ by more than the energy resolution.
\\
\indent The fit accounts for systematic uncertainties such as cut acceptance and the number of \isotope[125]{I} events by including them as fit parameter constraints. Additional systematics have to be considered when converting the observed number $N_\text{$2\upnu$ECEC}$ into a half-life. The \isotope[124]{Xe} isotopic abundance in XENON1T has been measured underground with a residual gas analyser (RGA) with a systematic uncertainty of $1.5\,\%$. The resulting abundance is $\eta = (9.94 \pm 0.14_\text{stat}\pm 0.15_\text{sys})\times 10^{-4}\;\frac{\text{mol}}{\text{mol}}$, which is $4\,\%$ larger than the natural abundance of $\eta = (9.52 \pm 0.03)\times 10^{-4}\;\frac{\text{mol}}{\text{mol}}$ \cite{deLaeter:2003abu}. The acceptance of the data selection criteria between \SI{55}{\keV} and \SI{75}{\keV} is constant within the uncertainties at $\epsilon = 0.967 \pm 0.007_\text{stat} \pm 0.033_\text{sys}$. The additional systematic uncertainty accounts for the fact that for a few data selection criteria only a lower limit on the acceptance was measurable.
The finite resolution of the position reconstruction in XENON1T leads to an uncertainty on the fiducial mass. This was quantified by contrasting the mass fraction, derived from the fiducial volume geometry and LXe density of \SI{2.862}{\gram/ \cubic\cm} at \SI{-96.1}{\celsius} \cite{nist:web}, with the fraction of \isotope[83\text{m}]{Kr} events in the fiducial volume. With this, the fiducial mass is $m = (1502 \pm 9_\text{sys})\;\text{kg}$. The half-life is then calculated as
\begin{align}
    T_{1/2}^{2\upnu\text{ECEC}} = \ln(2)\frac{\epsilon\,  \eta\, N_A\, m \,  t}{M_\text{Xe}\, N_{2\upnu\text{ECEC}}},\nonumber
\end{align}
where $M_\text{Xe}$ is the mean molar mass of xenon, $N_A$ is Avogadro's constant, and $t$ is the live-time of the measurement. The resulting half-life for the K-shell double electron capture of \isotope[124]{Xe} is $T_{1/2}^{2\upnu\text{ECEC}}=\nolinebreak(1.8\pm\nolinebreak 0.5_\text{stat}\pm\nolinebreak 0.1_\text{sys})\times 10^{22}\;\text{y}$. This is the longest half-life ever measured directly. Indications for a similarly-long half-life for $2\upnu$ECEC decay were reported for \isotope[78]{Kr} \cite{PhysRevC.96.065502}. Within the uncertainties the half-lives are equally long, but the uncertainty of our new result for \isotope[124]{Xe} is about two times smaller. Furthermore, the result is compatible with the lower limit from XMASS \cite{Abe:2018gyq}.\\
\indent This first direct observation of $2\upnu$ECEC in \isotope[124]{Xe} illustrates how xenon-based Dark Matter search experiments, with their ever-growing target masses and simultaneously decreasing background levels, are becoming relevant for other rare event searches and neutrino physics. It sets the stage for $0\upnu$ECEC searches that can complement double $\upbeta$-decay experiments in the hunt for the Majorana neutrino. Related processes involving the emission of one or two positrons ($2\upnu$EC$\upbeta^{+}$, $2\upnu\upbeta^{+}\upbeta^{+}$, $0\upnu$EC$\upbeta^{+}$, $0\upnu\upbeta^{+}\upbeta^{+}$) in \isotope[124]{Xe} might also exhibit interesting experimental signatures. The next generation detectors XENONnT \cite{aprile15c}, LZ \cite{Mount:2017qzi} and PandaX-4T \cite{Zhang:2018xdp} are already around the corner and will be able to probe these yet unobserved decays with unprecedented sensitivity. 

\cleardoublepage
\section*{Methods}
\begin{figure}
    \centering
    \includegraphics[width=0.45\textwidth]{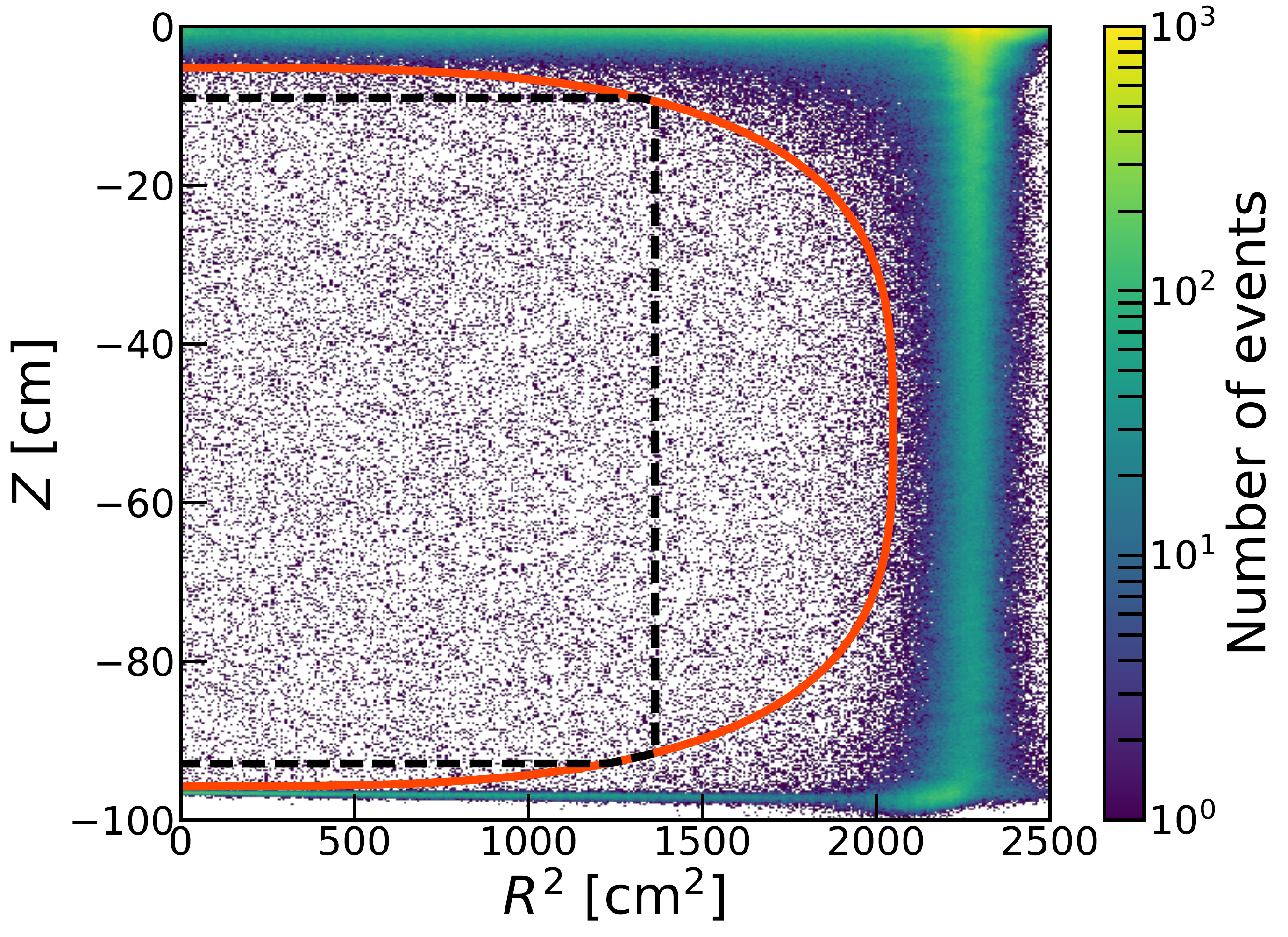}
    \caption{Spatial distribution in interaction depth $z$ vs. squared radius $R^2$ of events in a \SI{80}{\keV}-\SI{140}{\keV} window. High density areas correspond to the edges of the TPC where the majority of external $\upbeta$- and $\upgamma$-radiation is absorbed. The 1502 kg fiducial volume is indicated by the solid red line. The further segmentation into an inner (1.0 t) and outer (0.5 t) volume is marked by the black dashed line.}
    \label{fig:fiducial}
\end{figure}

\begin{figure}
    \centering
    \includegraphics[width=0.45\textwidth]{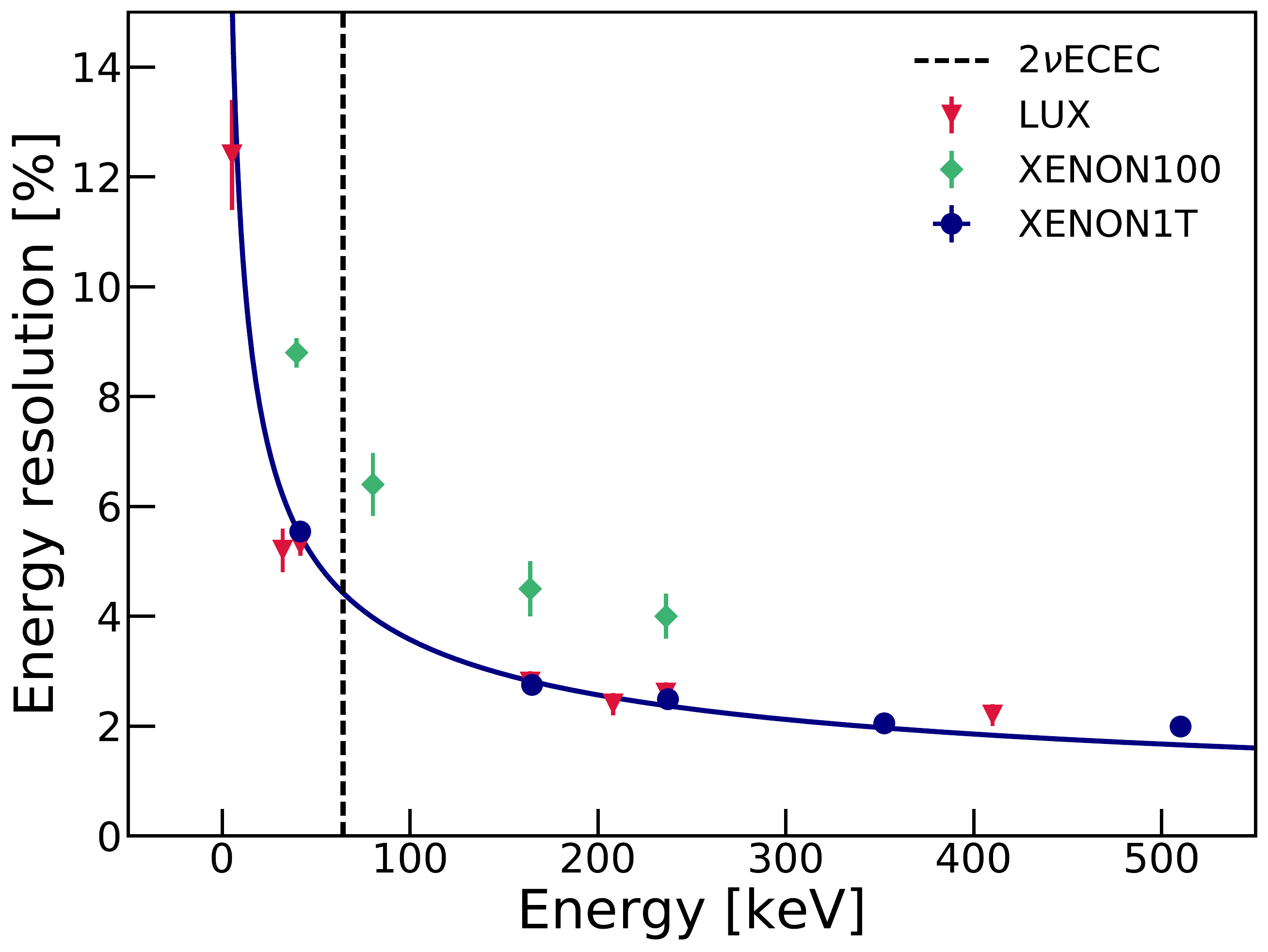}
    \caption{Energy resolution of low energy mono-energetic lines for selected liquid xenon Dark Matter experiments \cite{PhysRevD.95.012008, APRILE2012573} and the XENON1T detector in the 1.5 tonne fiducial mass. The relative resolution is defined as $\sigma_\text{E}/\mu_\text{E}$ of the Gaussian lines and fitted using a phenomenological function (solid blue line). For XENON1T the data points are \isotope[83\text{m}]{Kr} (41.5 keV), \isotope[131\text{m}]{Xe} (163.9 keV), \isotope[129\text{m}]{Xe} (236.2 keV), \isotope[214]{Pb} (351.9 keV) and $^{208}$Tl (510.8 keV). Only statistical uncertainties are shown for XENON1T which are too small to be visible. The energy of the $2\upnu$ECEC peak is indicated by the black dashed line.}
    \label{fig:resolution}
\end{figure}

\indent\textbf{Selection of the fiducial mass.} Since the $2\upnu$ECEC signal is proportional to the number of \isotope[124]{Xe} nuclei, it grows linearly with the xenon mass of the volume selected for the analysis $m_\text{volume}$. The ability to distinguish signal events from background depends on the background uncertainty $\Delta N_\text{background}$. For a counting experiment, the uncertainty on the number of background events $N_\text{background}$ is of Poissonian nature, so one has $\Delta N_\text{background}=\sqrt{N_\text{background}}$. The discovery sensitivity in a detector volume $S_\text{vol}$ is then proportional to the xenon mass in the selected volume divided by the background uncertainty:
\begin{align}
    S_\text{vol} \propto \frac{m_\text{volume}}{\sqrt{N_\text{background}}}.
\end{align} 
The $S_\text{vol}$ parameter was optimised using an automated algorithm that tests both cylindrical and superellipsoidal volumes. A 1502-kg-mass superellipsoid was found to give the optimal sensitivity. As the signal region was blinded, the optimisation was carried out in an energy sideband from \SI{80}{\keV} to \SI{140}{\keV}. For the fit of Monte Carlo simulations to the measured energy spectrum and consistency checks, the volume was segmented into an inner and outer volume (as indicated in Fig. \ref{fig:fiducial}). Intrinsic background sources mixed with the xenon, solar neutrinos, and $2\upnu$ECEC signal are expected to show the same activity in both volumes. However, the contribution from material backgrounds is strongest near the outer surface of the volumes. Fitting both volumes simultaneously gives a more robust fit and higher sensitivity than a single monolithic volume.
\\
\indent \textbf{Energy calibration and resolution.} Mono-energetic lines from the $\upgamma$-decays of four different isotopes are used for the energy calibration of the XENON1T detector. \isotope[83\text{m}]{Kr} is a gaseous calibration source that is homogeneously distributed inside the detector \cite{Manalaysay:2009yq}. The isomer undergoes a multi-step decay that is highly converted and deposits \SI{41.5}{\keV} inside the detector. This represents the lowest mono-energetic calibration point. The metastable \isotope[131\text{m}]{Xe} (163.9 keV) and \isotope[129\text{m}]{Xe} (236.2 keV) are neutron-activated during calibration campaigns and decay with half-lives of \SI{11.86}{\day} and \SI{8.88}{\day}, respectively. The \SI{1173.2}{\keV} and \SI{1332.5}{\keV} transitions of \isotope[60]{Co}, which is present in the stainless steel detector components such as the cryostat, are the highest energy calibration lines. Only energy depositions where the total energy of the $\upgamma$-transition is deposited in a single resolvable interaction within the detector are taken into account, i.e. the full absorption peak. The S1 and S2 signals from these interactions are then used to determine the yields of light and charge per unit energy for each source. The two quantities are anti-correlated \cite{Aprile:2007qd}, resulting in:
\begin{align}
    E = W\cdot\left(\frac{cS1}{g_1} + \frac{cS2_b}{g_2}\right)
\end{align}
at a given energy $E$. Here, $W=  (13.7\pm 0.2)\;\text{eV}$  \cite{Szydagis:2011tk} is the average energy needed to generate measurable quanta in LXe (S1 photons or S2 electrons), and $cS1$ and $cS2_b$ are the measured S1 and S2 signals corrected for detector-effects. S1 is corrected for the spatially dependent S1 light collection efficiency, whereas S2 is corrected for the spatial dependencies of both the charge amplification and the S2 light collection efficiency. The subscript on $cS2_b$ identifies the S2 signal seen by the bottom PMT array that is used for energy reconstruction in order to minimise the impact of signal saturation and non-uniformity due to single inactive PMTs in the top array. A fit to the measured data points gives the detector-specific calibration parameters $g_1$ and $g_2$. The calibration procedure is carried out in ten slices along the central axis of the cylindrical detector, in order to account for the depth dependence of $g_1(z)$ and $g_2(z)$ for the energy reconstruction.\\
\indent The energy resolution is determined from the reconstructed spectrum by fitting Gaussian functions with the mean $\mu_\text{E}$ and standard deviation $\sigma_\text{E}$ to mono-energetic peaks of the calibration sources (\isotope[83\text{m}]{Kr}, \isotope[131\text{m}]{Xe}, \isotope[129\text{m}]{Xe}) and radioactive isotopes in the TPC materials (\isotope[214]{Pb}, \isotope[208]{Tl}) up to \SI{510.8}{\keV}. The relative resolution is then given by $\sigma_\text{E}/\mu_\text{E}$ for each peak. The data points are finally fitted with a phenomenological function
\begin{align}
    \frac{\sigma_\text{E}}{\mu_\text{E}} = \frac{a}{\sqrt{E}} + b,
\end{align}
which gives an energy resolution of $4.1\,\%$ at the $2\upnu$ECEC energy (Fig. \ref{fig:resolution}).
\\
\indent \textbf{Iodine removal.} Thermal neutrons can be captured by \isotope[124]{Xe} producing \isotope[125]{Xe}: 
\begin{align}
    \isotope[124]{Xe} + \text{n} &\rightarrow \isotope[125]{Xe} + \upgamma.
\end{align}
\isotope[125]{Xe} decays to \isotope[125]{I} via electron capture with a half-life of \SI{16.9}{\hour}:
\begin{align}
   \isotope[125]{Xe} \xrightarrow[\text{EC}]{16.9\;\text{h}}\, &\isotope[125]{I}^* + \upnu_\text{e},\nonumber \\
   &\isotope[125]{I}^* \xrightarrow[\text{}]{<1\;\text{ns}}\, \isotope[125]{I} + \upgamma + \text{X}.
\end{align}
The X-rays and Auger electrons from the atomic relaxation after the electron capture are denoted by X. Iodine also undergoes electron capture to \isotope[125]{Te} with a \SI{59.4}{\day} half-life:
\begin{align}
    \isotope[125]{I} \xrightarrow[\text{EC}]{59.4\;\text{d}}\, &\isotope[125]{Te}^* + \upnu_\text{e},\nonumber \\
    &\isotope[125]{Te}^* \xrightarrow[\text{}]{1.48\;\text{ns}}\, \isotope[125]{Te} + \upgamma + \text{X}.
\end{align}
Both decays populate short-lived excited nuclear states of \isotope[125]{I} and \isotope[125]{Te} and the signals from the $\upgamma$-transitions are merged with the atomic relaxation signals following the electron capture. The Te K-shell X-ray, which has a branching ratio of $87.5\,\%$, is merged with a \SI{35.5}{\keV} nuclear transition. This is problematic because it makes a Gaussian line centred around \SI{67.3}{\keV}, which is about $1\sigma$ away from the \SI{64.3}{\keV} expected for $2\upnu$ECEC.\\
\indent Two significant mechanisms leading to the presence of \isotope[125]{I} in the detector have been identified: artificial activation during calibration campaigns by neutrons from the deuterium-deuterium fusion neutron generator or the $^{241}$AmBe source, and activation outside of the water shield by environmental thermal neutrons. As the decay rate of \isotope[125]{Xe} can be monitored during and after calibration campaigns, one can predict the decay rate of its iodine daughter. For environmental neutrons, flux measurements at LNGS are used to estimate the activation. These estimates are cross-checked with the \isotope[125]{Xe} decay peaks in the data. In both post-AmBe and post-neutron generator data, fewer iodine decays than expected from the decay of the mother isotope \isotope[125]{Xe} were found. This is attributed to the removal of \isotope[125]{I} during the continuous purification of the detector's xenon inventory by circulation over hot zirconium getters. Due to the blinding of the signal region that contains the \isotope[125]{I} peak, the long-term behaviour of the removal could only be assessed after unblinding.
\begin{figure}
    \centering
    \includegraphics[width=0.45\textwidth]{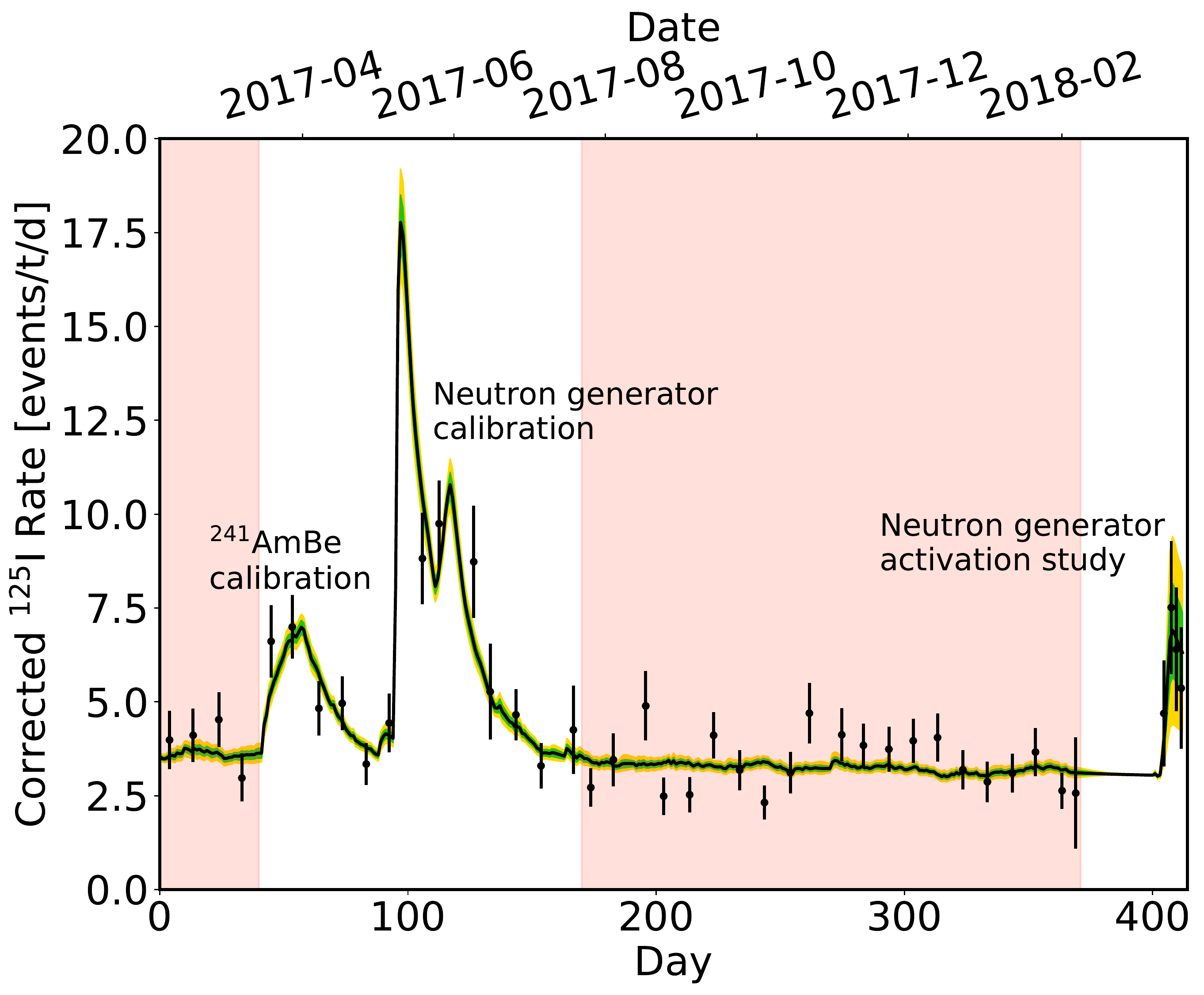}
    \caption{Fit of the \isotope[125]{I} time evolution model to data in a $2\sigma$ interval around the \isotope[125]{I} peak mean in 10-day bins. Periods with increased \isotope[125]{I} decay rate are attributed to artificial activations from neutron calibrations, equipment tests, and a dedicated activation study. The decrease of the rate to the background level corresponds to an effective iodine decay constant $\tau = 9.1\;\text{d}$. The best fit is shown as the solid black line. The green (yellow) bands mark the $1\sigma$ ($2\sigma$) model uncertainties resulting from the Poisson uncertainties of the \isotope[125]{Xe} data underlying the model. The data selection for the $2\nu$ECEC search, where the decay rate has returned to the background level, is indicated in pale red.}
    \label{fig:i125}
\end{figure}\\
\indent As every \isotope[125]{Xe} decay in the detector leads to the presence of an \isotope[125]{I} nucleus, a model for the expected iodine decay rate from artificial activation is constructed by integrating the background-subtracted \isotope[125]{Xe} rate over time in one-day steps. The data is then convolved with the effective decay constant $\tau$ and fitted with a free amplitude and linear background to the measured \isotope[125]{I} rate evolution in a $2\sigma$ interval around the peak (\SI{61.7}{\keV} to \SI{72.9}{\keV}). An effective \isotope[125]{I} decay constant of $\tau = (9.1\pm 2.6)\;\text{d}$ was found, which is in agreement with an expected decay constant from completely efficient getter removal.\\
\indent Since the model is constructed directly from data, the uncertainties from the \isotope[125]{Xe} rates are propagated by introducing artificial Poisson fluctuations to the data points. An \isotope[125]{I} model is made for each variation of the \isotope[125]{Xe} data and fitted to the \isotope[125]{I} rate evolution. The best fit to the \isotope[125]{I} rate over time in 10-day bins and the uncertainty band derived from an ensemble of 1,000 fits are shown in Fig \ref{fig:i125}. Different binnings between 1 and 14 days have been tested for consistency with $\chi^2$ and log-likelihood fits.\\
\indent An integration of each model over the actual data taking periods yields an expected number of \isotope[125]{I} decays $N_\text{\isotope[125]{I},art}$. The ensemble distribution of $N_\text{\isotope[125]{I},art}$ allows to extract both a central value and uncertainties. Now, only data sets with a decay rate at the non-activated background level are selected for the $2\upnu$ECEC search. The final data selection is shown in Fig. \ref{fig:i125}. For the spectral fit of the remaining $177.7$ live days we constrain the number of expected iodine events from artificial activation $N_\text{\isotope[125]{I},art}$ using the model. We also constrain the radiogenic component $N_\text{\isotope[125]{I},rad}$ taking into account the effective decay constant $\tau$.
\begin{figure}
    \centering
    \includegraphics[width=0.45\textwidth]{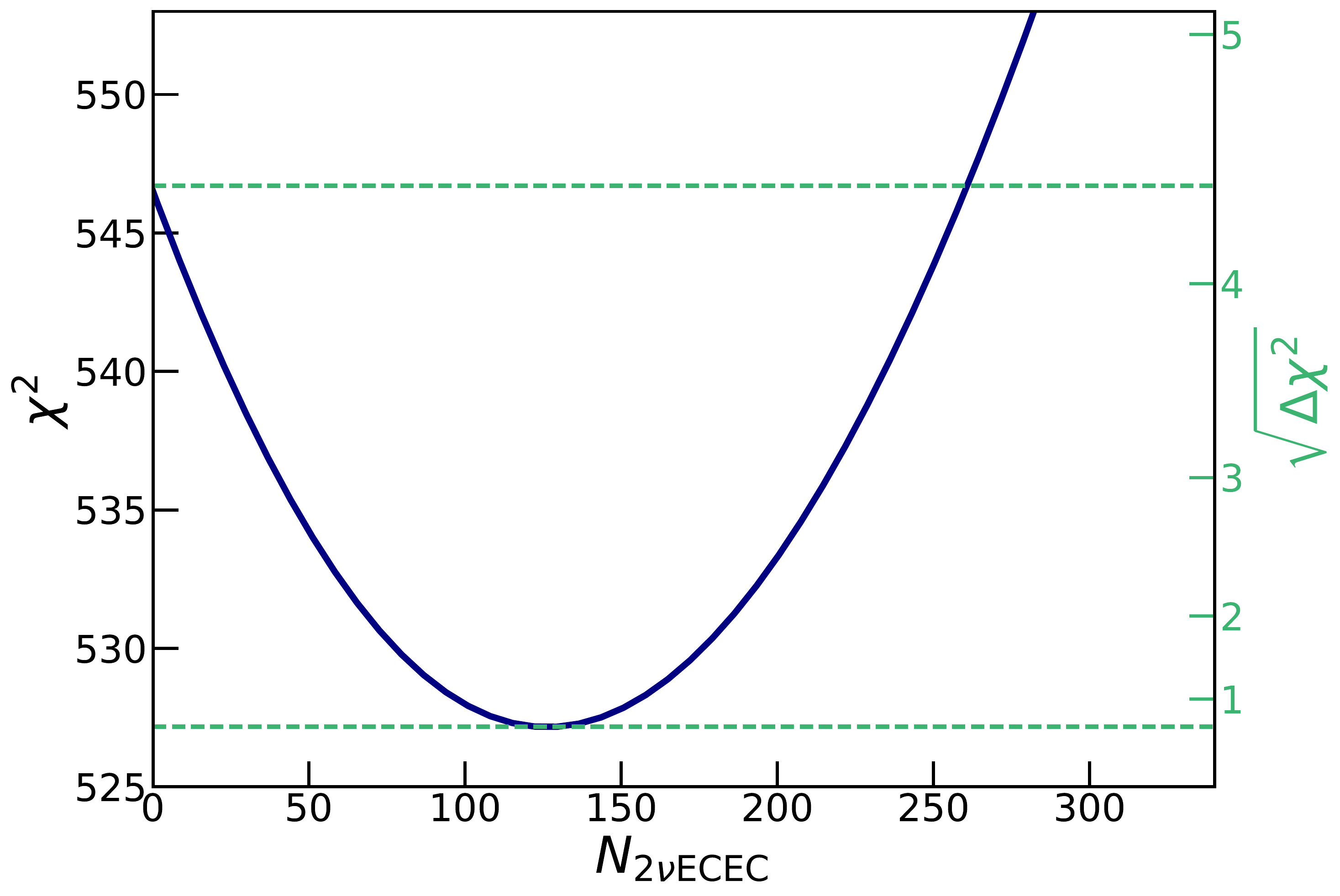}
    \caption{$\chi^2$ curve for the number of measured $2\upnu$ECEC events. Comparing the best fit value of $N_\text{$2\upnu$ECEC}= 126\;\text{events}$ to a null result one obtains $\sqrt{\Delta \chi^2} = 4.4$.}
    \label{fig:chisquare}
\end{figure}\\
\indent \textbf{Fit method.} The data is fitted with all known background sources, either simulated or modelled as Gaussian peaks, and the $2\upnu$ECEC peak. The scaling parameters of the simulated Monte Carlo spectra and the properties of the Gaussian peaks are the fit parameters in a $\chi^2$ minimisation
\begin{align}
\chi^{2}_\text{combined}(\vec p) = \sum_\text{i} \frac{(R_\text{i}-f(E_\text{i},\,\vec p))^2}{(\Delta R_\text{i})^2},
\end{align}
where $R_\text{i}$ is the measured event rate in the energy bin $E_\text{i}$ and $f(E_\text{i},\,\vec p)$ is the background fit function. At energies below \SI{100}{\keV}, low statistics of simulated backgrounds from detector construction materials require an interpolation of the simulated spectra in order to avoid over-fitting. As the main background contribution from materials in this energy region are single Compton scatters from $\upgamma$-rays in the sensitive volume, a featureless spectrum is expected. Thus, the sum of the material contributions is linearly interpolated up to \SI{100}{\keV}. This gives
\begin{align}
    f(E_\text{i},\,\vec p) &= \left[\sum_\text{k}^\text{materials} p_\text{k} R_\text{k}(E_\text{i})\right]_\text{interpolated $< 100$ keV}\nonumber\\
    &+ \sum_\text{l}^\text{intrinsic} p_\text{l} R_\text{l}(E_\text{i})\nonumber \\
    &+ \sum_\text{m}^\text{Gaussians} \text{Gaussian}_\text{m}\,(\vec p_\text{m},\,E_\text{i}),
\end{align}
where the sums correspond to the interpolated material component, the intrinsic sources plus solar neutrinos and the Gaussian peaks with the fit parameters $p_\text{k,l,m} \in \vec p$. Knowledge from external measurements, such as material screening \cite{Aprile:2017scr}, \isotope[85]{Kr} concentration measurements \cite{Aprile:2018prl} and elemental abundances have been incorporated into the fit function and are constrained using terms of the form
\begin{align}
    \text{constraint}_\text{j} = \frac{(\text{parameter}_\text{j}-\text{expectation}_\text{j})^2}{\text{uncertainty}_\text{j}^2}.
\end{align}
A deviation of the fit parameter by $n \times \sigma$ from the expectation will thus increase the value of the $\chi^2$ function by $n^2$. The Gaussian signal peak has been constrained in the fit as well given the prior information on the expected position and width. Moreover, systematic uncertainties from the cut acceptance and fiducial mass are addressed by including these as constrained fit parameters in the fit function. As the fit is carried out in an inner and outer detector volume, each of the two volumes has its own $\chi^2$-function with distinct parameters for the respective fiducial masses $\vec V$ and cut acceptances $\vec \kappa$. The energy reconstruction was found to agree within the uncertainties. The full $\chi^2$ function can then be written as:
\begin{align}
\chi^{2}_\text{combined}(\vec p,\, \vec V,\, \vec\kappa)&= \chi^{2}_\text{inner}(\vec p,\, V_\text{inner},\,\kappa_\text{inner})\nonumber \\
&+\chi^{2}_\text{outer}(\vec p,\, V_\text{outer},\,\kappa_\text{outer})\nonumber  \\
&+ \text{constraint}_{\vec p}\nonumber\\
&+ \text{constraint}_\text{V}\nonumber \\ 
&+ \text{constraint}_\kappa.
\end{align}
More details of the background modelling will be discussed in a future publication.\\
\indent \textbf{Fit result.} The $\chi^2$ curve for the number of observed $2\upnu$ECEC events is shown in Fig. \ref{fig:chisquare}. The $4.4\sigma$ significance is derived from the $\Delta \chi^2$ between the best fit and a null result along the curve.\\
\indent \textbf{Data availability.}
The data that support the findings of this study is available from the corresponding authors upon reasonable request.

\section*{Acknowledgements}
We thank Javier Men\'endez for sharing his expertise in the theory of double $\upbeta$-decay. We gratefully acknowledge support from the National Science Foundation, Swiss National Science Foundation, German Ministry for Education and Research, Max Planck Gesellschaft, Deutsche Forschungsgemeinschaft, Netherlands Organisation for Scientific Research (NWO), NLeSC, Weizmann Institute of Science, I-CORE, Pazy-Vatat, Initial Training Network Invisibles (Marie Curie Actions, PITNGA-2011-289442), Fundacao para a Ciencia e a Tecnologia, Region des Pays de la Loire, Knut and Alice Wallenberg Foundation, Kavli Foundation, Abeloe Graduate Fellowship, and Istituto Nazionale di Fisica Nucleare. Data processing is performed using infrastructures from the Open Science Grid and European Grid Initiative. We are grateful to Laboratori Nazionali del Gran Sasso for hosting and supporting the XENON project.
\section*{Author Contributions}
The XENON1T detector was designed and constructed by the XENON Collaboration. Operation, data processing, calibration, Monte Carlo simulations of the detector and of theoretical models, and data analyses were performed by a large number of XENON Collaboration members, who also discussed and approved the scientific results. The analysis presented here was performed by a large number of XENON Collaboration members. The paper was written by A.Fi. and C.Wi. It was reviewed and edited by the collaboration and all authors approved the final version of the manuscript. 
\section*{Author Information}
Author Information Reprints and permissions information is available at www.nature.com/reprints. The authors declare no competing financial interests. Readers are welcome to comment on the online version of the paper. Correspondence and requests for materials should be addressed either to C.Wi. (c.wittweg@uni-muenster.de), to A.Fi. (a.fieguth@uni-muenster.de) or to the XENON collaboration (xenon@lngs.infn.it).


\end{document}